\documentclass[12pt]{iopart}
\usepackage{iopams}
\usepackage{graphicx}
\usepackage{placeins}
\usepackage{caption}
\usepackage{subcaption}
\usepackage{cite}

\usepackage{color}

\begin{document}

``This is the Accepted Manuscript version of an article accepted for publication in New Journal of Physics (NJP).  IOP Publishing Ltd is not responsible for any errors or omissions in this version of the manuscript or any version derived from it.  The Version of Record is available online at https://doi.org/10.1088/1367-2630/ac7d6e."

\title[]{Surface Plasmon-Driven Electron and Proton Acceleration without Grating Coupling}

\author{J.~Sarma$^1$, A.~McIlvenny$^1$, N.~Das$^2$, M.~Borghesi$^1$, and A.~Macchi$^{3,4}$ }

\address{$^1$Centre for Plasma Physics, The Queen’s University of Belfast, University Road BT71NN, Belfast, United Kingdom
\newline$^2$Tezpur University, Tezpur, India
\newline $^3$National Institute of Optics, National Research Council (CNR/INO), Adriano Gozzini laboratory, Pisa, Italy
\newline $^4$Enrico Fermi Department of Physics, University of Pisa, Pisa, Italy}
\ead{jsarma01@qub.ac.uk and m.borghesi@qub.ac.uk}
\vspace{10pt}

\date{\today}

\begin{abstract}
  Surface plasmon (SP) excitation in intense laser interaction with solid target can be exploited for enhancing secondary emissions, in particular efficient acceleration of high charge electron bunches. Previous studies have mostly used grating coupling to allow SP excitation, which requires stringent laser contrast conditions to preserve the structural integrity of the target. Here we show via simulations that efficient SP electron acceleration for currently available short pulse lasers can occur in a flat foil irradiated at parallel or grazing incidence ($\sim 5^\circ$ with the target surface) without a surface modulation. In turn, the accelerated electrons can be effective for generating proton beams with narrow spectra peaked at $>$100~MeV energies for currently available laser drivers.
 
\end{abstract}

\section{Introduction}
\label{Introduction}
Surface plasma waves or surface plasmon polaritons, hereby referred to as surface plasmons (SP) for brevity, are electromagnetic modes localized at, and propagating along a sharp interface between, e.g., vacuum and a conducting medium (metal or plasma). Exciting SP by a laser pulse can lead to strong field enhancement at the interface, which has numerous applications in plasmonics \cite{maier2007plasmonics}. In the context of high-intensity laser interactions with solid targets, a number of experiments provided evidence of SP excitation mainly via enhancement of secondary emissions such as XUV photons (either as incoherent radiation \cite{gauthierSPIE95,kahalyPRL08} or coherent high harmonics \cite{cantono2018extreme}), protons or ions \cite{bagchiPoP12,ceccotti2013evidence}, and electrons \cite{huPoP10,mishimaRLE15,fedeli2016electron,cantono2018extensive,zhuHPLSE20}. The latter, in particular, can be accelerated along the surface by ``surfing'' the SP, similarly to what happens in a plasma wakefield. These electrons are characterized by high energy (with respect to the quiver energy in the laser field \cite{fedeli2016electron}) and high total charge (up to several hundreds of picoCoulombs \cite{cantono2018extensive}), with simulations showing that the electrons form multiple bunches with few femtosecond duration \cite{cantono2018extreme}.

The above mentioned experiments used grating targets, i.e. targets engraved with a shallow periodic modulation to allow the coupling of the SP with the laser pulse. This is considered necessary since for a plane electromagnetic EM wave impinging at an angle $\theta$ over a flat plasma-vacuum interface the phase matching conditions cannot be satisfied. In fact, phase matching for linear excitation requires the wavevector component along the surface ($x$ direction for definiteness) of the EM wave and the SP to be equal, i.e. $k_{{\rm EM},x} = k_{{\rm SP},x}$. 
If $\omega$ is the frequency of the waves, we have
\begin{equation}
  k_{{\rm EM},x} = \frac{\omega}{c}\sin{\theta}, \qquad
  k_{{\rm SP},x} = \frac{\omega}{c} \left( \frac{\epsilon(\omega)}{\epsilon(\omega)+1}\right)^{1/2} = 
  \frac{\omega}{c}\left( \frac{1-\omega_p^2/\omega^2}{2-\omega_p^2/\omega^2}\right)^{1/2} \; ,
\end{equation}
where we assumed a simple plasma dielectric function $\varepsilon=\varepsilon(\omega)=1-\omega_p^2/\omega^2$ with $\omega_p$ the plasma frequency, and $\omega_p/\omega>\sqrt{2}$ as a necessary condition for SP existence. Since $k_{{\rm EM},x} < \omega/c < k_{{\rm SP},x}$, the condition $k_{{\rm EM},x} = k_{{\rm SP},x}$ is impossible. However, in a medium with periodicity in the $x$-direction, due to the Floquet-Bloch theorem the phase matching condition becomes
\begin{equation}
  k_{{\rm EM},x} = k_{{\rm SP},x} + nq \; ,
  \label{eq:FB}
\end{equation}
with $n$ an integer and $q$ the wavevector of the periodic modulation. This allow the resonant SP excitation at those values of $\theta$ for which Eq.(\ref{eq:FB}) is satisfied. Actually one does not need an infinite grating, but it is sufficient to have a local surface modulation extended over a few wavelengths, which allows the excitation of SP with a tightly focused pulse as it is typical of high intensity experiments.

While in ordinary plasmonics one may use prism-based schemes \cite{maier2007plasmonics} as an alternative to grating coupling, such approach is not suitable for ultrashort, intense laser pulses because of strong dispersion and nonlinear effects in the prism material. Moreover, in both contexts the grating coupling has the disadvantage that an SP propagating along the grating surface will lose energy due to radiative scattering (the inverse of the excitation process). The energy loss might be compensated by engraving the target only in the laser spot region \cite{marini2021ultrashort}, but this is challenging at high intensities due to the limited pointing stability of high power femtosecond systems. Using such systems also require efficient contrast enhancement strategies (such as double plasma mirrors\cite{dromey2004plasmamirror,thaury2007plasmamirror,levy2007doublePlasmaMirror}) to prevent prepulse damage of the shallow grating.

Recently, a new geometry apparently suitable for SP acceleration of electrons without a grating has been proposed by Shen \textit{et al}\cite{pukhov}. In the basic proposed scheme, named ``peeler acceleration'', the laser is incident on the short edge of a thin foil, and parallel to the foil surface. This geometry may allow the laser pulse to excite an SP because, since the foil has a finite length along the $x$ direction, translational symmetry along $x$ is broken and thus the matching of wavevectors is not required. Nevertheless, in the limit $\omega_p/\omega \gg 1$ (which is well satisfied at solid densities) one has $k_{{\rm SP},x} \to k_{{\rm EM},x}$, having posed $\theta=90^{\circ}$ for parallel incidence, with the laser propagating along $x$ like the SP, and at almost the same phase velocity. The simulations in Ref.\cite{pukhov} show that SP electron acceleration is rather efficient in the proposed geometry, which in turn also allows for exploitation of the energized electrons to drive proton acceleration. In fact, the total charge of the electrons which are peeled by the laser and then accelerated by the SP can exceed those of protons placed at the opposite short edge of the foil target. Thanks to the excess space charge, the sheath electric field formed at the edge is smooth and allows to produce a narrow peaked proton spectrum. (Incidentally, a similar albeit less prominent effect can be noticed in simulations shown in Ref.\cite{cristoforetti2020laser} where an array of parallel foils is simulated in two dimensional (2D) Cartesian geometry.) The proposed scheme looks very promising for applications of proton acceleration and calls for experimental verification. However, a successful implementation may require to address technical issues such as the limited pointing stability which may make it  difficult to hit a micron-thick target on the short edge with high precision.  

\begin{figure}[t]
     \centering
     \includegraphics[scale=0.25]{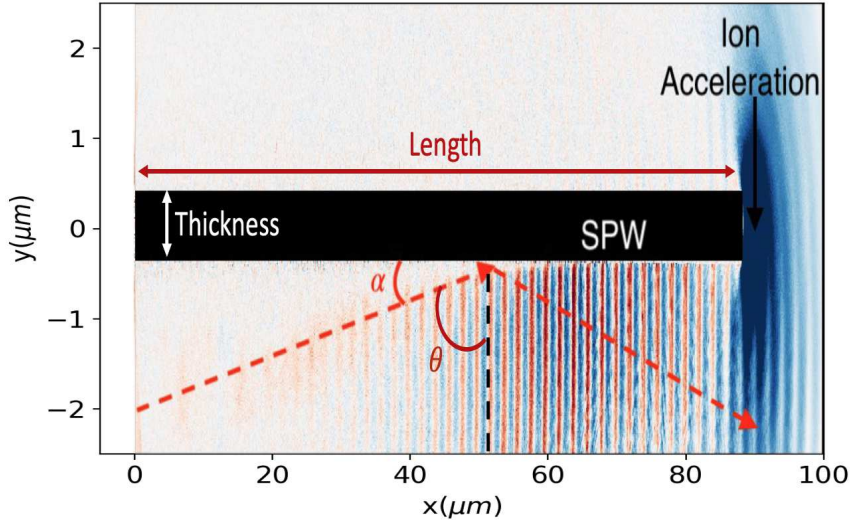}
     \caption{Schematic of the interaction, showing the target irradiated at grazing incidence and the region of ion acceleration. The red dashed lines represent the incident and reflected laser. A contourplot of the $E_x$ field components is also shown.}
     \label{SPW_fig}
\end{figure}

In anticipation of experiments, this paper reports on a simulation study in configurations similar to the proposed ``peeler'' scheme. Two main findings of experimental relevance are apparent from our simulations. 
First, the acceleration is effective also for grazing incidence of the laser pulse, which may relax issues related to laser alignment and pointing. 
Second, when the laser pulse is incident in the direction parallel to the target, the highest electron and proton energies may be obtained with a shifted laser spot, i.e. with the laser axis not lying in the target midplane ($y=0$ in the schematic of Fig.\ref{SPW_fig}).

\FloatBarrier

\section{Simulation set-up}
Two-dimensional (2D) particle-in-cell (PIC) simulations have been performed using the open source code EPOCH \cite{arber2015contemporary}. A 2D simulation grid with resolution $\Delta x = 10$ nm and $\Delta y = 2\Delta x$. The $y$-range of the grid was [-22$\mu$m, 10$\mu$m] in all simulations while the $x$-range [0 $\mu$m, 96 $\mu$m] was varied between different simulations. Open boundary conditions (for EM and particles) have been set on all sides of the simulation box. The laser pulse propagates along the positive $x$ direction. For all the simulations shown, the laser pulse has a central wavelength $\lambda=0.8~\mu$m  and a Gaussian profile along the transverse ($\perp$) direction. For propagation in the $x$ direction (which would correspond to parallel incidence, i.e. $\theta=0$, see Fig.\ref{SPW_fig}) the transverse field profile is 
\begin{equation}
E_y(x,y)=E_0 \exp{(-y^2/w^2)}\exp{(ik_{\rm EM}x)} \; ,
\end{equation}
with $w=4.9\lambda=3.9~\mu$m which corresponds to a focal spot diameter of $1.665w=6.5~\mu$m full width at half maximum (FWHM). The pulse duration is $35$~fs, also FWHM. Note that both width and duration are referred to the field profile; for the intensity profile, the corresponding FWHM values are shorter by a factor $\sqrt{2}$.  

The peak intensity $I$ of the pulse has been varied berween $3.4 \times10^{19}$~W/cm$^2$ and $I=7.8 \times 10^{20}$~W/cm$^2$ which correspond to values of the dimensionless intensity parameter $a_0=4$ and $a_0=19$, respectively, being $a_0=0.85\sqrt{I\lambda^2/10^{18}\rm{W cm}^{-2}\mu\rm{m}^2}$. Notice that for the lowest value of $I$ the laser parameters are very close to those of Refs.\cite{fedeli2016electron,cantono2018extensive}, allowing us to compare our results with previous grating-based experiments and simulations.

The target left edge is placed at $x=0$ and the length along $x$ has been varied in simulations, while the target is always $1\lambda=0.8\mu$m thick along $y$ and centered at $y=0$. The electron density is  $n_e = 100n_c$ where $n_c=(\pi m_ec^2/e^2\lambda^2)$ is the critical or cut-off density corresponding to the laser wavelength ($n_c\simeq1.7 \times 10^{21}$~cm$^{-3}$ for $\lambda=0.8~\mu$m). The target is composed of Au ions except for a  CH layer of $0.32~\mu$m thickness added at the right edge of the target. The numbers of macroparticles used per cell are 200 for electrons, C and H ions and 100 for Au ions.

Fig.\ref{SPW_fig} shows a schematic of the interaction, including the definition of the incidence angle $\theta$ and its complementary grazing (or glancing) angle $\alpha=90^{\circ}-\theta$, and of the target length and thickness. The laser is incident from the $y<0$ region and propagates from left to right. Ion acceleration occurs at the right short edge of the target. The figure includes a snapshot of the $E_x$ field generated at grazing incidence.

\section{Grazing incidence}
\label{sec:grazing}

\subsection{Theory}

Before showing simulation results, we discuss how grazing incidence on a flat surface may enable to excite and sustain a SP suitable for efficient electron acceleration. As mentioned above, the phase matching condition strictly holds for a plane EM wave of infinite extension and duration. A non-resonant excitation of an SP may be possible in transient conditions and taking finite width effects into account. Suppose that an SP is excited by the leading edge of the incident laser pulse (conditions for efficient coupling will be discussed below). The laser and SP wavefronts will travel with different phase velocities,
\begin{equation}
    v_{{\rm EM},x} = \frac{\omega}{k_{{\rm EM},x}} \; , \qquad v_{{\rm SP},x} = \frac{\omega}{k_{{\rm SP},x}} \; .
\end{equation}
After propagating over a distance $L$, the wavefronts will accumulate 
a phase difference 
\begin{equation}
    \Delta \phi = (k_{{\rm EM},x} - k_{{\rm SP},x})L = \left(\frac{1}{v_{{\rm EM},x}}-\frac{1}{v_{{\rm SP},x}}\right)\omega L \; .
\end{equation}
After a distance $L_d$ such that $\Delta \phi =\pi$, i.e.
\begin{equation}
    L_d=\frac{\pi}{k_{{\rm SP},x}-k_{{\rm EM},x}} \; , 
\end{equation}
the waves will cancel each other due to destructive interference. Still a SP wave will be sustained over a length less than $L_d$, which will be larger for smaller differences between the phase velocities. The latter become very close in the conditions of grazing incidence $(\sin{\theta}\to 1)$ and high density ($n_e/n_c=\omega^2_p/\omega^2\gg 1$).
To first order both in $\alpha=\pi/2-\theta$ (in radians) and $n_c/n_e$, we obtain
\begin{equation}
  L_d \simeq \frac{\lambda}{\alpha^2+n_c/n_e} \; . 
  \label{dephasing}
\end{equation}
Note that this expression is similar to Eq.(2) in Ref.\cite{pukhov} but with the laser diffraction angle $\theta_d=\lambda/(\pi w)$ replaced by $\alpha$. 
Using for example $\theta = 85^{\circ}$ (so that $\alpha=8.7 \times 10^{-2}$)
and $n_e/n_c = 100$, Eq.(\ref{dephasing}) yields $L_d \simeq 57\lambda$. Note that the effective laser spot size along the target surface will be $\Delta x_s \simeq w/\cos{\theta}=w/\sin\alpha$ with $w$ the laser waist, hence for tightly focused pulses $\Delta x_s$ will be typically comparable to $L_d$. This implies that the laser pulse may sustain the SP all over the dephasing length.

Now considering the acceleration
of electrons trapped in the SP field, the typical value of the final energy is \cite{fedeli2016electron}
${\cal E} \simeq m_ec^2 a_{{\rm SP}}(n_e/n_c)=(2/\pi)(eE_{{\rm SP}}\lambda)(n_e/n_c)$
where $E_{{\rm SP}}=\mbox{max}(E_{{\rm SP},x})$ is the peak value of the SP longitudinal field component and $a_{{\rm SP}}=(eE_{{\rm SP}}/m_e\omega c)$. This corresponds to an acceleration length $L_a \simeq {\cal E}/(eE_{{\rm SP},x}) = (\lambda/\pi)(n_e/n_c)$, which is independent of the SP field amplitude. Thus, the large value of $L_d$ at grazing incidence may yield values of $L_a$ sufficient to obtain the maximum allowable electron energy. Note that the comparison between simulation and experiment in Ref.\cite{fedeli2016electron} suggested that $L_a$ was limited to the laser spot width in those conditions.

Of course, the intensity on target will decrease with angle as $I(\theta)=I(0)\cos\theta=I(0)\sin\alpha$. However, with respect to the scenario of grating coupling (which for focused pulses typically require $\theta \lesssim 45^{\circ}$ in order to illuminate more than one grating period), the intensity decrease at grazing incidence may be compensated by three additional factors: i) the electron energy is proportional to the SP electric field, hence it has a slow $\sim\cos^{1/2}\theta=\sin^{1/2}\alpha$ scaling with the angle; ii) there is no SP energy loss due to radiative scattering by the grating; iii) the regime may remain far from the onset of a weaker scaling with $a_0$ which was previously observed in numerical simulations at $a_0 \gtrsim 10$  \cite{fedeli2016electron}. Here we may note an analogy with laser wakefield acceleration where a quasi-linear regime may be more favorable for acceleration than a highly nonlinear one \cite{berkeley}.

In addition, grazing incidence may allow a more efficient coupling between the laser pulse and SP by allowing the laser field in the skin layer to have the same instantaneous direction as the SP field. This geometrical condition is optimal to excite the SP since the electrons will be driven along the same trajectory that they perform in the SP field. Let us represent the laser pulse by a $P$-polarized plane wave with amplitude $E_0$, incident from the $y<0$ halfspace at an angle of incidence $\theta$. The electric field components of the EM wave at $y=0^{-}$, i.e. just below the $y=0$ interface with the plasma medium are (omitting the temporal dependence)
\begin{eqnarray}
  E_{{\rm EM},x}(y=0^{-})&=&-E_0(1-r)\cos\theta \; , \\
  E_{{\rm EM},y}(y=0^{-})&=&+E_0(1+r)\sin\theta \; ,
\end{eqnarray}
where $r$ is the Fresnel coefficient
\begin{equation}
  r=\frac{\varepsilon\cos\theta-(\varepsilon-\sin^2\theta)^{1/2}}
         {\varepsilon\cos\theta+(\varepsilon-\sin^2\theta)^{1/2}} \; .
\end{equation}  
Immediately inside the skin layer $(y=0^{+}$) we obtain
\begin{eqnarray}
  E_{{\rm EM},x}(y=0^{+})&=&E_{{\rm EM},x}(y=0^{-}) \; , \\
  E_{{\rm EM},y}(y=0^{+})&=&\frac{1}{\varepsilon} E_{{\rm EM},y}(y=0^{-}) \; ,
\end{eqnarray}
so that
\begin{eqnarray}
  \left.\frac{E_{{\rm EM},x}}{E_{{\rm EM},y}}\right|_{y=0^{+}}&=&
  -\varepsilon\frac{1-r}{1+r}\frac{\cos\theta}{\sin\theta}
  =-\frac{(\varepsilon-\sin^2\theta)^{1/2}}{\sin\theta} \nonumber \\
  &=&-i\frac{(\omega_p^2/\omega^2-1+\sin^2\theta)^{1/2}}{\sin\theta}
  \; . \label{eq:REM}
\end{eqnarray}
For the SP field components we have (see e.g. \cite{macchi2018surface})
\begin{equation}
  \left.\frac{E_{{\rm SP},x}}{E_{{\rm SP},y}}\right|_{y=0^{+}}=
  -i \left(\frac{\omega_p^2}{\omega^2}-1\right)^{1/2} \; . \label{eq:RSP}
\end{equation}
Optimal coupling between the EM and SP waves would require the expressions (\ref{eq:REM}) and (\ref{eq:RSP})  to be equal, which is algebraically impossible. However, the difference gets very small in the limit $\sin\theta\to 1$.  This shows that grazing or, ideally, parallel incidence favors the coupling. Note that this simple argument neglects the energy transfer from the EM wave to the SP. Since the energy flux into the $y>0$ region goes to zero anyway when $\sin\theta\to 1$, sustaining the SP wave growth will require $\theta<\pi/2$, i.e. a non-vanishing grazing angle.

\begin{figure}[t]
  
  \includegraphics[width=\textwidth]{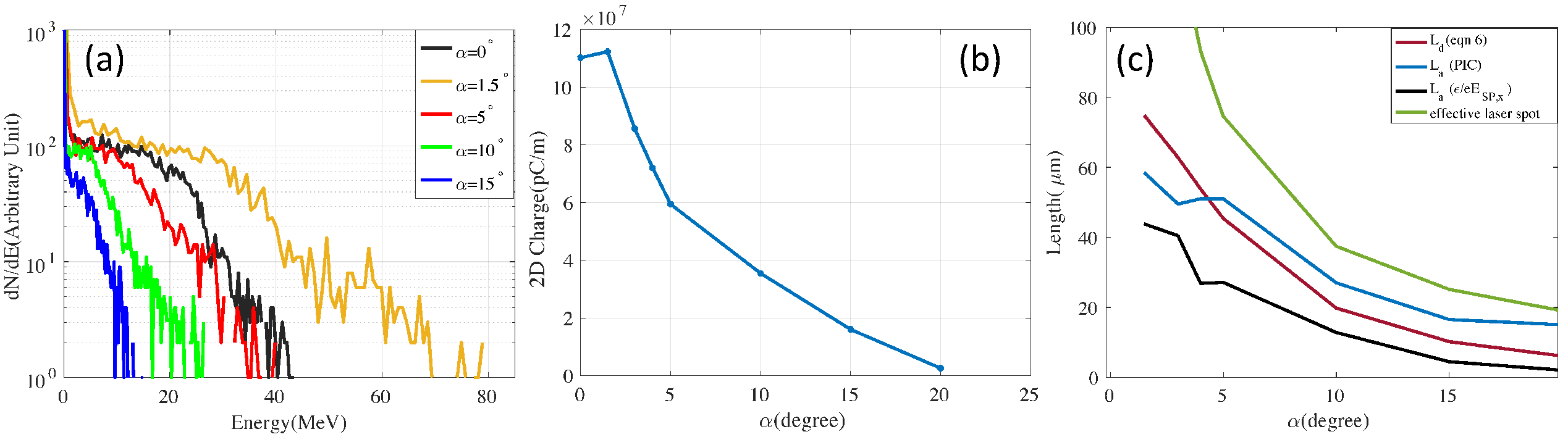}
  \caption{(a) Energy spectra of electrons, (b) accelerated electron charge, and (c) estimates of the acceleration length (see text for details) as a function of the grazing angle $\alpha$, compared to the dephasing length $L_d$ [Eq.(\ref{dephasing})] and the spot size FWHM$/\sin\alpha$. In these simulations, $I=3.4 \times10^{19}$~W/cm$^2$ ($a_0=4$) and the target length is $90~\mu$m for a) and b) and $200~\mu$m for c).}
\label{with_angle}
\end{figure} 
\FloatBarrier

\subsection{Simulation results}
Indeed, efficient acceleration at grazing incidence is observed in our simulations.
Figure \ref{with_angle}~(a) shows the electrons energy spectra for different laser incidence angles corresponding to the grazing angle range $1.5^{\circ}<\alpha<20^{\circ}$. For a comparison, the case of parallel incidence ($\alpha=0^{\circ}$)  is included; this corresponds to the basic geometry studied in Ref.\cite{pukhov} albeit with the laser centred along the target midplane as opposed to the target edge (this will further discussed in Section~\ref{sec:parallel} below). 
Only electrons from the laser-irradiated surface and with ejection angle $\phi=\arctan(p_y/p_x)$ in the range $1^{\circ}<\phi<10^{\circ}$ and $p_x>0$ (being $p_y$ and $p_x$ the electron momentum components along $y$ and $x$, respectively) have been considered. The spectra are obtained at time $t=300$~fs, just before the electron bunch reaches the target rear edge and at this time the energy gain is almost over. Here $t=0$~fs is the time when the peak of the pulse enters the simulation box. The electron energy increasing with decreasing $\alpha$ down to 1.5$^{\circ}$ for which the cut-off value is 70~MeV, much higher than what obtained at parallel incidence.  
For a further comparison, experimental and simulation results reported in Ref.\cite{cantono2018extensive} for grating targets show electron energies up to 20~MeV for the same laser parameters.

\begin{figure}[t]
  \includegraphics[width=\textwidth]{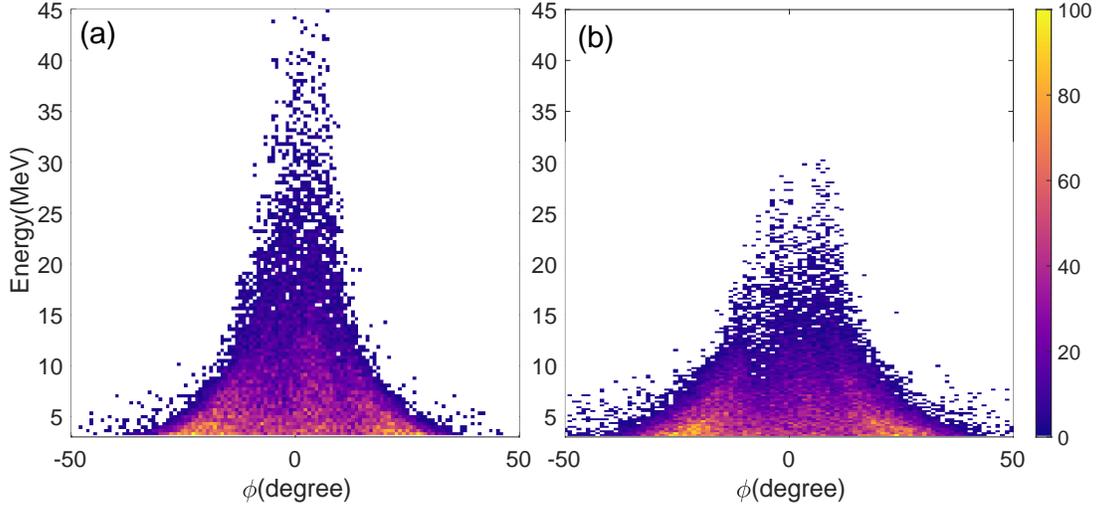}
  \caption{Energy-angle distribution of electrons for grazing incidence (a) $\alpha = 5^\circ$ (b) $\alpha = 10^\circ$. The colour scale represents the relative number of electrons.}
\label{angdistr_angle}
\end{figure} 
\FloatBarrier

Along with the energy, the number of accelerated electrons also increases with decreasing $\alpha$. Figure \ref{with_angle}~(b) shows the total charge in pC/m as a function of $\alpha$. To roughly estimate the amount of charge that would be accelerated in a real 3D geometry, we multiply the linear charge density by the pulse FWHM diameter ($6.5~\mu$m), since in 3D the interaction would be extended for such a length along the $z$ direction. This yields, for $\alpha=1.5^{\circ}$, a total charge of $\sim 780$~pC which is higher than all measurements reported in Ref.\cite{cantono2018extensive} including the best case of a blazed grating for which $\sim 660$~pC were measured.

In order to estimate the acceleration length ($L_a$), a longer target of $200\mu$m length has been used to ensure that the target is long enough for electrons to gain the maximum energy before reaching the target rear edge. For this aim, the temporal change in the energy spectra of electrons has been monitored. After a certain amount of time $\tau_a$ the energy does not increase anymore and the spectrum remains unchanged. Considering the electrons velocity to be near luminal, $L_a \simeq c\tau_a$ has been estimated and shown as the blue curve in Fig.\ref{with_angle}~c). As an independent estimate, we evaluated the average longitudinal field $E_{{\rm SP}0}$ from the simulations and estimated $L_a \simeq U_{max}/eE_{{\rm SP}0}$, where $U_{max}$ is the maximum energy and $e$ is the electron charge: the results are shown as the black curve of Fig.\ref{with_angle}~c). Although the two estimates differ quantitatively (not surprising because of the roughness of the methods), both curves are reasonably close to the dephasing length $L_d$ [Eq.(\ref{dephasing})] and much below the spot width.

The energy-angle distribution of electrons (Fig.\ref{angdistr_angle}) shows that the higher the energy the closer the electron direction to the tangent, which is in agreement with the simple picture of acceleration in a SP \cite{fedeli2016electron,macchi2018surface}. Consistently, the electron beam collimation is stronger for lower values of $\alpha$ for which higher energies are reached.

\begin{figure}[t]
  \includegraphics[width=\textwidth]{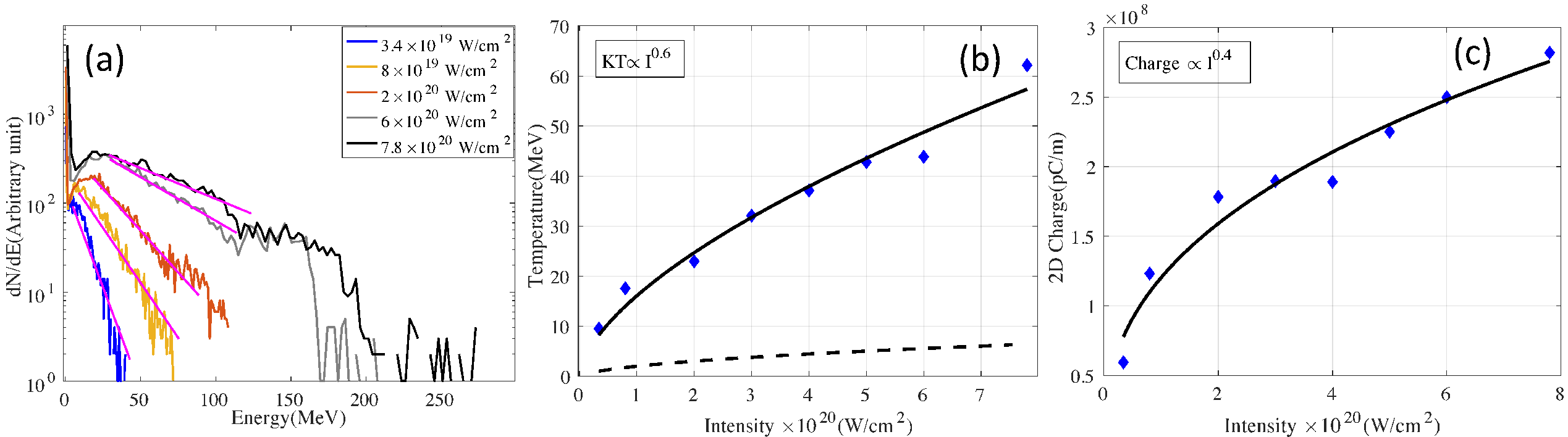}
  \caption{(a) Electron energy spectra, (b) electron temperature, $T_e$ (solid line) and Ponderomotive scaling of temperature, $T_{pond}$ (dashed line), and (c) total electronic charge as functions of laser intensity, at a grazing angle of $\alpha = 5^{\circ}$.}\label{energy_intensity}
     \label{Temp_vs_int} \label{Charge_int}
\end{figure}

\FloatBarrier

Fig.\ref{energy_intensity}~a) shows the variation of electron spectra with the laser intensity $I$ for the $\alpha=5^{\circ}$ case. At higher intensities the spectra become relatively flat near the cut-off energy, similar to previous observations in grating targets \cite{fedeli2016electron}, while the central part is well fitted by an exponential spectrum $\sim \exp(-{\cal E}/T_e)$ from which an effective temperature $T_e$ can be introduced. Fig.\ref{Temp_vs_int}~b) shows that $T_e$ scales as $\propto I^{0.6}$ with intensity and is much higher than the ``ponderomotive'' value \cite{wilks1992absorption} $T_{pond} = m_ec^2(\sqrt{(1+a_0^2/2)}-1)$.
For example, at $I=3.4 \times 10^{19}$~W/cm$^2$ we get $T_e \simeq 10~{\rm MeV} \gg T_{pond} \simeq 1~\rm{MeV}$. The total accelerated charge scales as $\propto I^{0.4}$ which is slower than the linear scaling observed in grating targets \cite{cantono2018extensive}. For the highest value of $I$, a total charge of $\simeq 1.9$~nC would be expected in 3D.

\begin{figure}[h]
     \centering
          \includegraphics[scale=0.3]{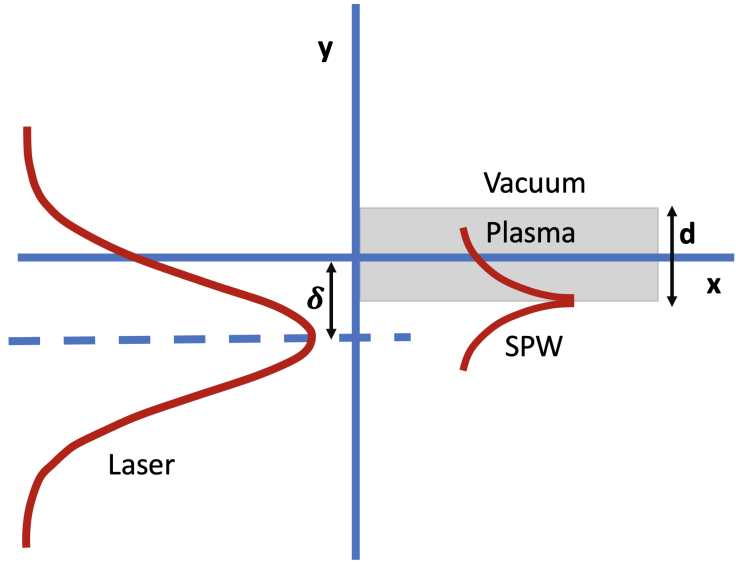}
     \caption{Schematic of Laser spot shift and SP propagation at the target surface. Here $y>0$ and $y<0$ are the plasma and vacuum region respectively. The laser peak is shifted by an amount $\delta$ from the plasma surface towards the vacuum.}
     \label{SP_L}
\end{figure}
\FloatBarrier

To conclude this section, we note that a previous numerical study of electron acceleration at grazing incidence was reported by Serebryakov et al. \cite{serebryakovPoP17}; while the PIC results show some similarities to ours, the mechanism is attributed to direct acceleration by the field resulting from the superposition of incident and reflected laser fields (yielding a superluminal phase front), without any mention of SP excitation.

\section{Parallel incidence}
\label{sec:parallel}

We now turn to simulations for parallel incidence ($\theta=90^{\circ}$ or $\alpha=0^{\circ}$ in which the laser pulse is sent parallel to the target surface, hitting the short edge of the target. This corresponds to the basic geometry proposed in Ref.\cite{pukhov} and further investigated in Ref.\cite{shenQE21}. Our primary aim was to test how electron and ion acceleration are affected by laser misalignment on a scale of a few wavelengths, which are likely to occur because of the limited pointing stability of high power systems.

\begin{figure}[h]
     \centering
     \includegraphics[width=0.48\textwidth]{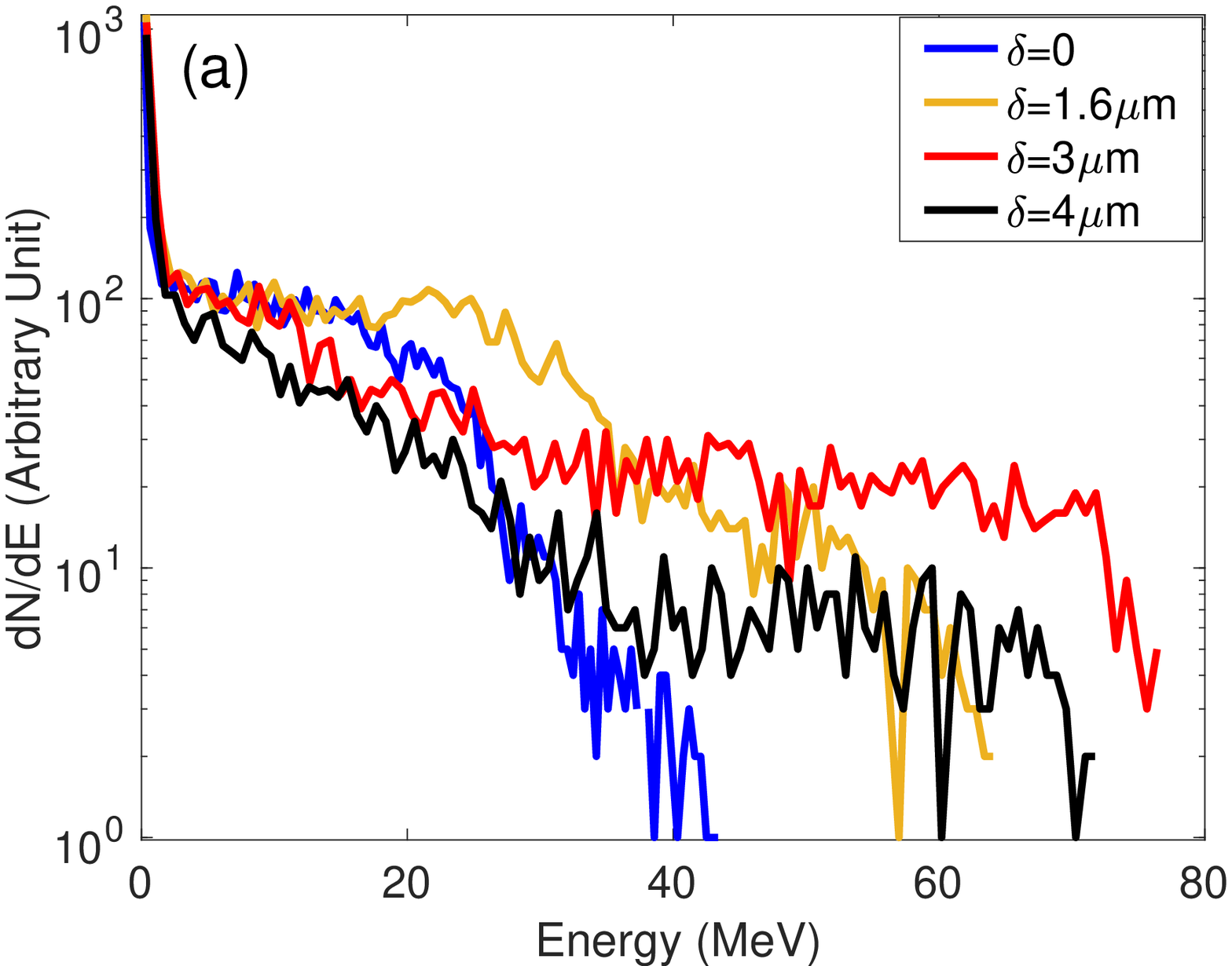}
     \includegraphics[width=0.48\textwidth]{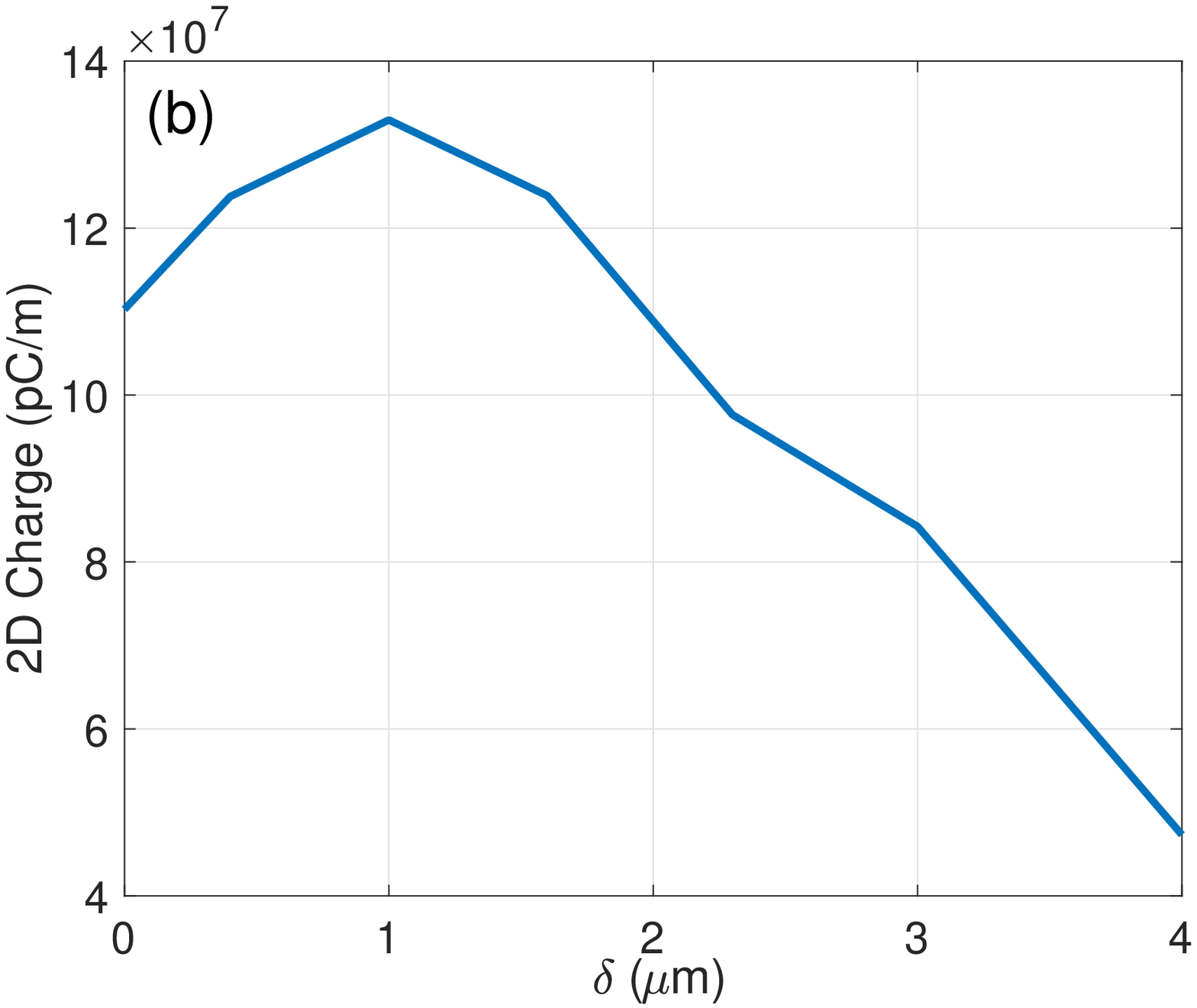}
     \caption{Parallel incidence of the laser pulse at low intensity ($I=3.4 \times10^{19}$~W/cm$^2$, $a_0=4$): (a) electron energy spectra and (b) total charge of accelerated electrons as a function of the focal spot shift (distance of the laser propagation axis from the target surface).}
\label{Energy_angle} \label{Electron_energy_misaln}
\end{figure}

\begin{figure}[t]
     \centering
     \includegraphics[width=0.48\textwidth]{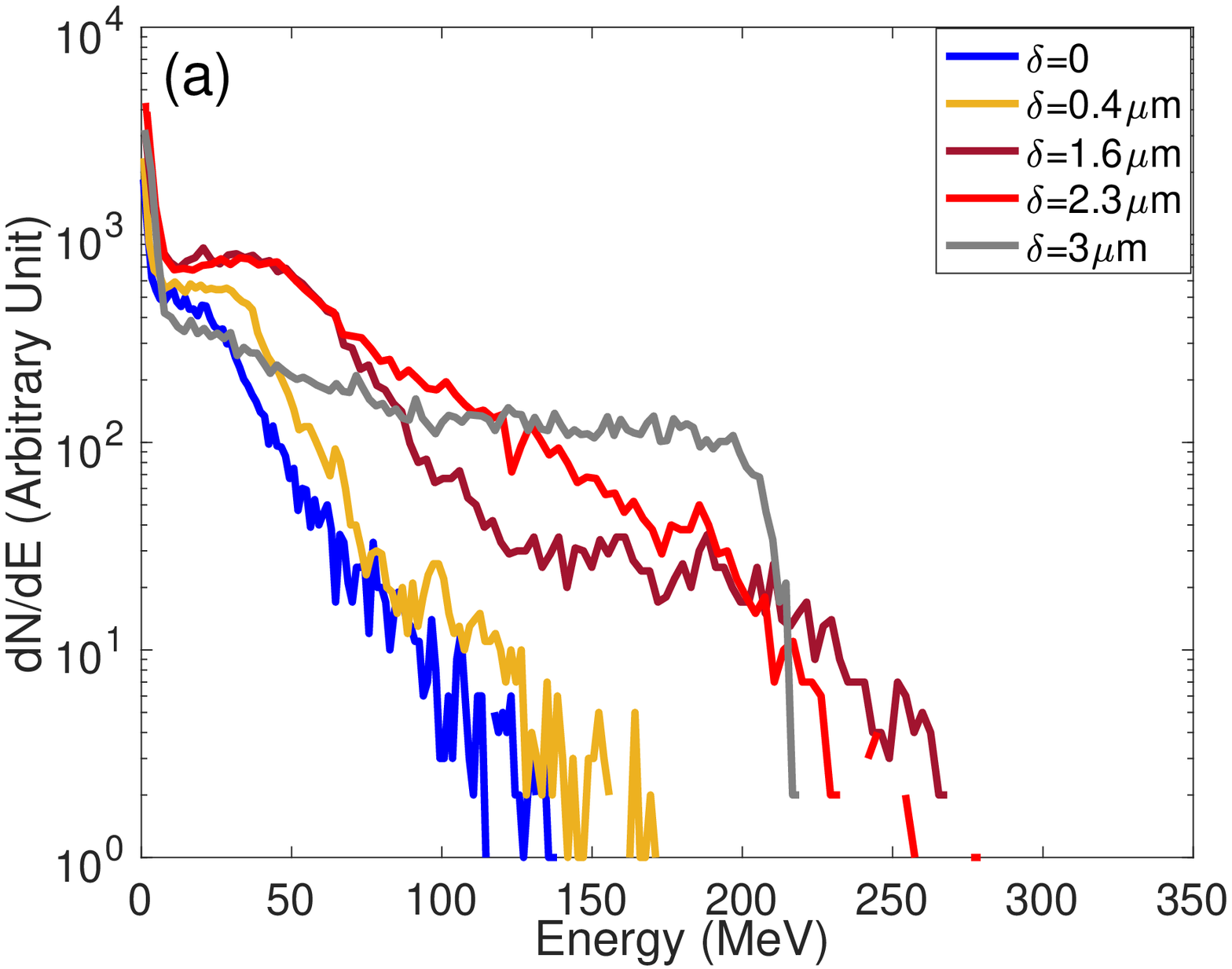}
     \includegraphics[width=0.48\textwidth]{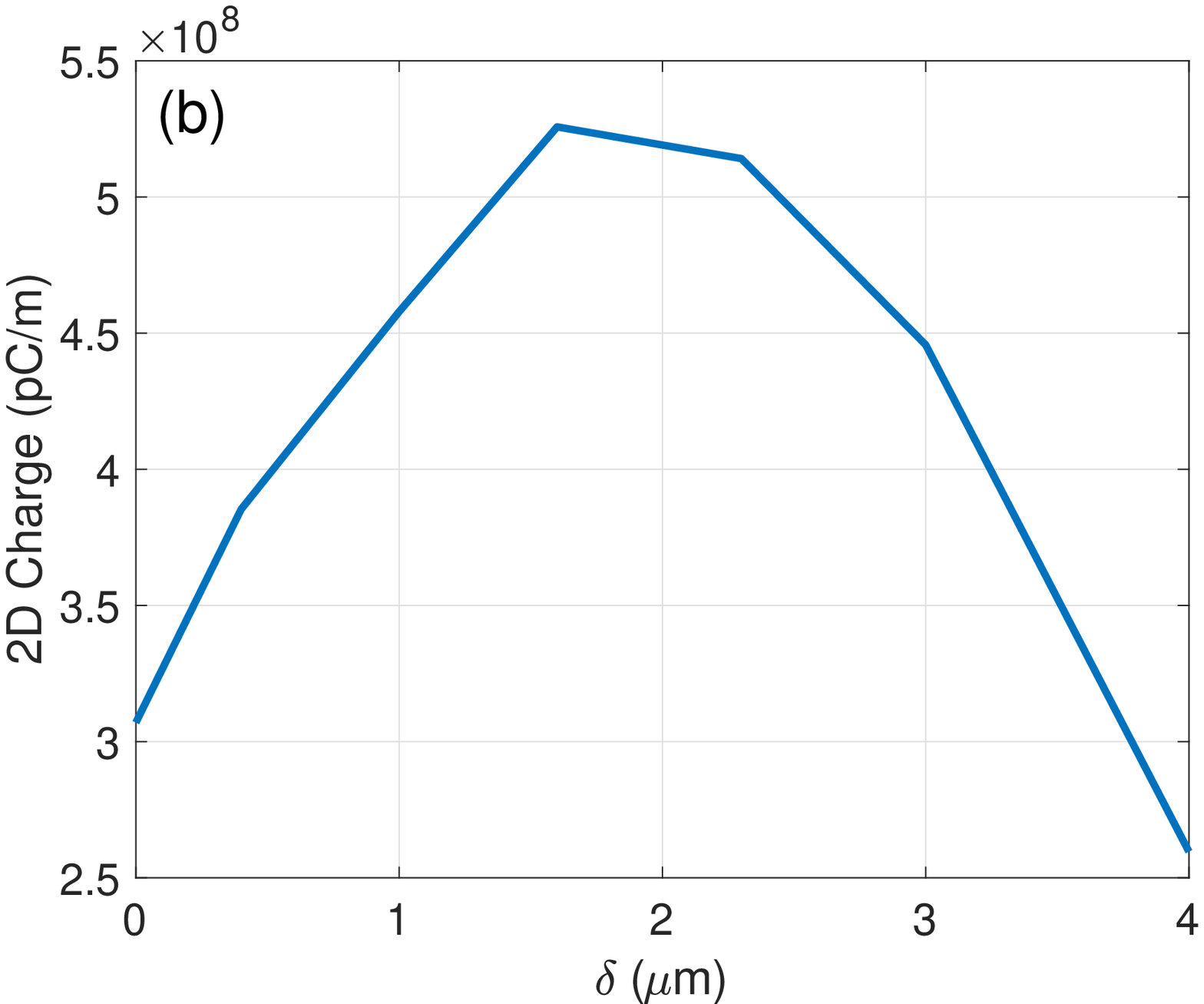}
\caption{Parallel incidence of the laser pulse at high intensity ($I=6.0 \times10^{20}$~W/cm$^2$, $a_0=16.6$): (a) electron energy spectra and (b) total charge of accelerated electrons as a function of the focal spot shift (distance of the laser propagation axis from the target surface).}
\label{Electron_energy_misaln-Gemini} \label{Energy_angle-Gemini}
\end{figure}

Indeed, our simulations show that a certain amount of focal spot shifting (as shown in the schematic of Fig.\ref{SP_L}), so that the laser pulse axis does not coincide with the target midplane, may lead to more efficient electron and proton acceleration by increasing the cut-off energy, the accelerated charge, and the beam collimation. Figs.\ref{Energy_angle} and \ref{Energy_angle-Gemini} show results for energy spectra and accelerated charge as the function of the shift $\delta$, for the lowest and highest intensities considered. In both cases, the cut-off energy and the total charge are maximized at a non-zero value of $\delta$.
Note that this effect could not be observed in Ref.\cite{shenQE21} because a transverse plane wave was used there.

\begin{figure}[t]
  \includegraphics[width=\textwidth]{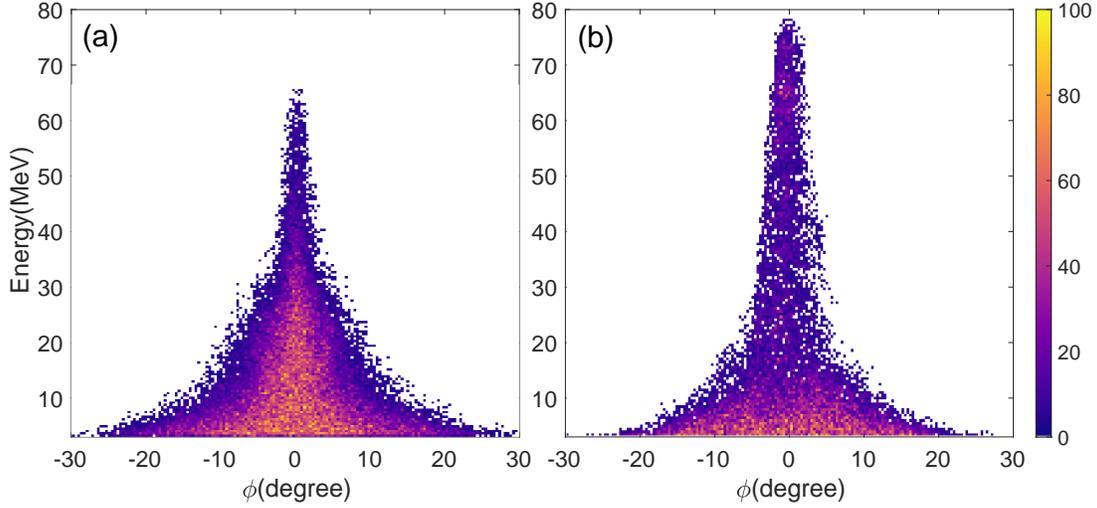}
  \caption{Energy-angle distribution of electrons for focal spot shift of (a) $\delta = 0$ $\mu$m (b) $\delta = 3$ $\mu$m. The colour scale represents the relative number of electrons. }
\label{angdistr_shift}
\end{figure} 
\FloatBarrier

For $I=3.4 \times10^{19}$~W/cm$^2$ (Fig.\ref{Energy_angle}) the highest energy is obtained for $\delta=3~\mu$m; the cut-off value of $\simeq 75$~MeV is nearly twice the value for $\delta=0$, i.e. when the laser is aligned with the target midplane ($y=0$). Increasing $\delta$ to $4~\mu$m only slightly decreases the cut-off energy, but significantly reduces the spectral density of electrons near the cut-off. The total accelerated charge at $\delta=1~\mu$m has a maximum value close to that obtained for grazing incidence at the same intensity (Fig.\ref{with_angle}~b). The energy-angle distribution (Fig.\ref{angdistr_shift}) shows that the most energetic electrons are strongly collimated. The ejection angle still decreases with energy and, consistently, the beam aperture is smaller for $\delta=3$ $\mu$m for which the highest energies are obtained. Also note that, with respect to the symmetrical interaction geometry of \cite{pukhov,shenQE21}, the energy-angle distribution remains symmetrical also for the configuration with the shifted pulse irradiating only one target side.

\FloatBarrier

For $I=6.0 \times 10^{20}$~W/cm$^2$ (Fig.\ref{Energy_angle-Gemini}), the maximum cut-off of $\simeq 250$~MeV is obtained for $\delta=1.6~\mu$m. This also corresponds to the maximum of the total charge, which is about twice the value obtained at a grazing incidence of $\alpha=5^{\circ}$ for the same intensity (Fig.\ref{Charge_int}~c), and would correspond to about $3.4$~nC in 3D.

The condition of exact parallel incidence, i.e. $\delta=0$, might not be optimal to sustain the growth of the SP because the EM energy flow across the surface becomes too small, as discussed above when analyzing the grazing incidence case (Section \ref{sec:grazing}). To go deeper into the comparison, we note that for a finite value of $\delta$ the ratio between the  $E_{{\rm EM},x}$ and $E_{{\rm EM},y}$ components at the surface of the target ($y=-d/2$) will be equal to the value for a plane wave incident at some angle $\theta_{\delta}$. 
For a Gaussian spatial profile the transverse electric field is $E_{{\rm EM},y}\simeq E_0 \exp{(-(y+\delta)^2/w^2)}\exp{(ik_{{\rm EM},x})}$. Using the zero divergence condition $\nabla \cdot {\bf E} = 0$, the longitudinal field component is obtained as $E_{{\rm EM},x} \simeq 2y/(ik_{{\rm EM},x}w^2)\exp{(-(y+\delta)^2/w^2)}\exp{(ik_{{\rm EM},x})}$. Thus, at $y=-d/2$ we obtain for the ratio between the $x$ and $y$ components 
\begin{equation}
    \left|\frac{E_{{\rm EM},x}}{E_{{\rm EM},y}}\right|_{y=\delta}\simeq \frac{(\delta-d/2)\lambda}{\pi w^2}
    \label{laser_field_ratio}
\end{equation}
Taking $\lambda=0.8\mu$m and $w=3.9\mu$m, the values of this ratio for $\delta=2.3~\mu$m (high $I$) and $3~\mu$m (low $I$) would be the same as for a $P$-polarized plane wave incident at angles $\theta_{\delta} \simeq 88.3^{\circ}$ and $\simeq 87.6^{\circ}$. The discussion in Section~\ref{sec:grazing} also indicates that the higher the fractional absorption $A$, the larger the deviation of the optimal angle from parallel incidence should be. From the simulations we obtain $A \simeq 28\%$ and 
$\simeq 40\%$ for high $I$ and low $I$, respectively, which qualitatively agrees with our expectations. 

Note that the energy and charge enhancement observed for $\delta>0$ cannot be explained by the amount of laser energy ``blocked'' by the target edge. From the simulation we estimate that the amount of laser energy reflected from the target edge is $\sim 11$\% and $\sim 6$\% for $\delta=0$ and $\delta=3$, respectively. This difference is not sufficient to explain the increase in performance indicating that the energy is being coupled more efficiently instead.

\section{Proton acceleration}

\begin{figure}[t]
     \centering
     \includegraphics[width=0.48\textwidth]{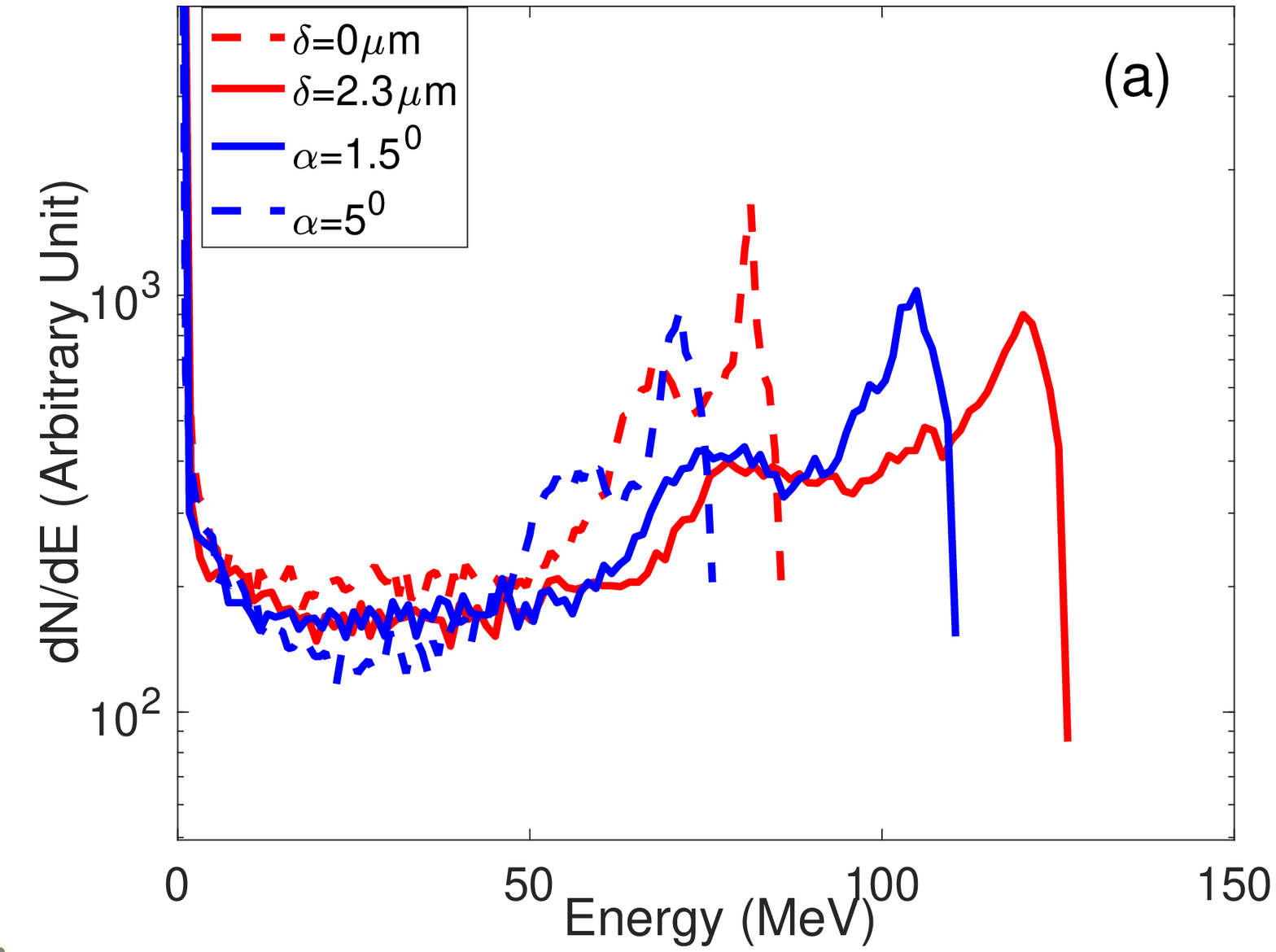}
     \includegraphics[width=0.48\textwidth]{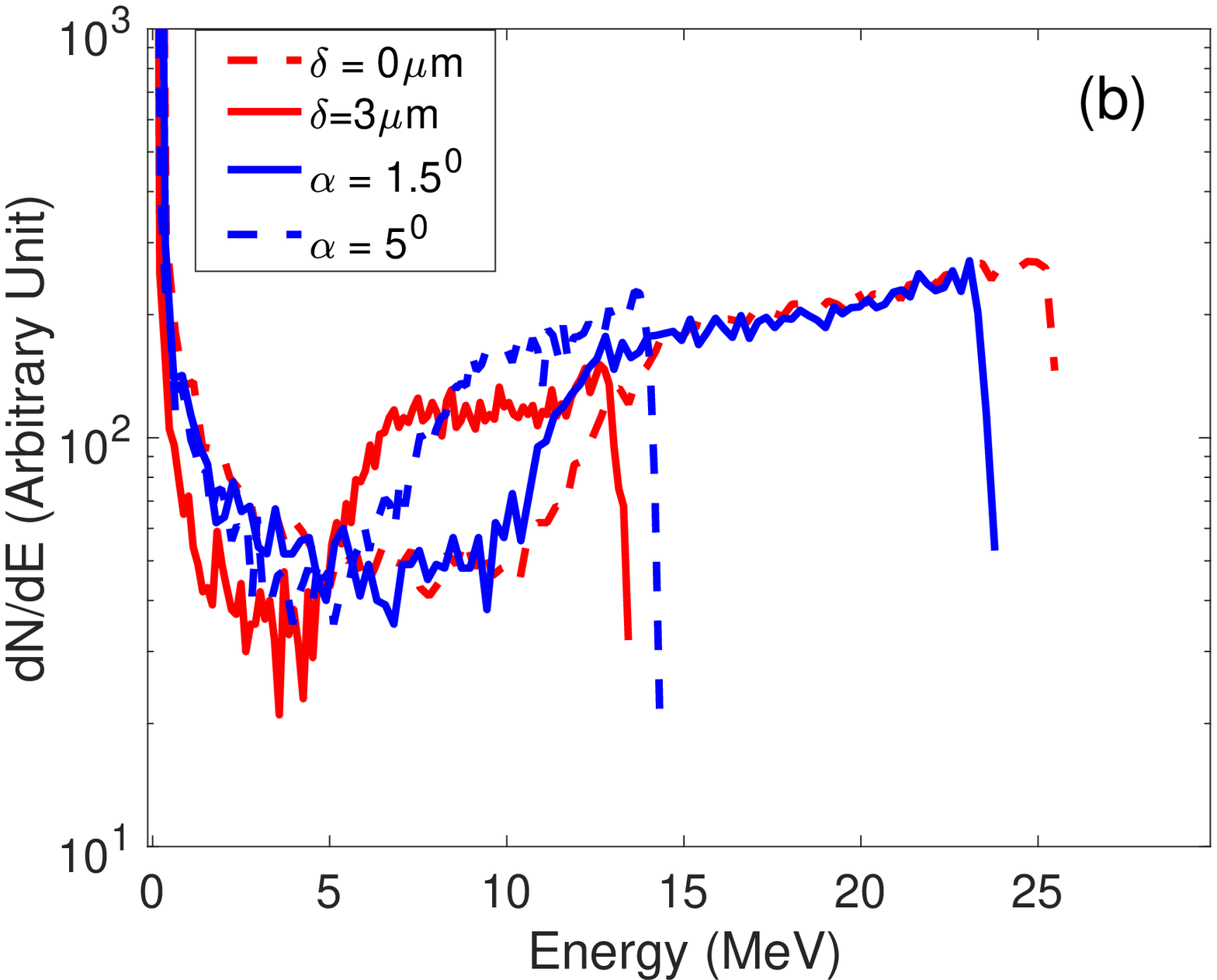}

  \caption{Energy spectra of protons accelerated from the rear edge of the target, for both grazing incidence at two different values of $\alpha$ and parallel incidence when the pulse propagation axis either coincides with the target midplane ($y=0$) or it shifted by $\delta$, and for two values of the laser intensity: a) $I=6 \times 10^{20}$ W/cm$^2$; b) $I=3.4 \times 10^{19}$~W/cm$^2$.}
     \label{Energy_proton_combined}
\end{figure}
\FloatBarrier

In this section we report on the acceleration of protons from the rear short edge of the target, as in the scheme proposed by Shen \textit{et al}.\cite{pukhov}, for both cases of grazing incidence and parallel incidence with a shifted laser pulse. The proton energies are measured at $t=450$fs (i.e. 150 fs after the electron spectra reported throughout this paper), after which no more significant acceleration happens and the energy becomes saturated. At the highest intensity ($I=6 \times 10^{20}$ W/cm$^2$) shown in Fig.\ref{Energy_proton_combined}~a), the highest proton energy cut-off of $\simeq 130$~MeV is obtained in the parallel configuration with a pulse shift of $\delta=2.3~\mu$m, which largely exceeds the value of $\simeq 85$~MeV for $\delta=0$. The grazing incidence cases with $\alpha=1.5^{\circ}$ and $\alpha=5^{\circ}$ yield cut-off values of $\simeq 110$~MeV and $\simeq 75$~MeV, respectively. For all the cases shown, the spectrum is strongly peaked at the cut-off with a second, broader peak at lower energy.

Note that the cut-off energy of $\simeq 130$~MeV exceeds the value of $\simeq 100$~MeV obtained in the 3D simulations of Shen et al.\cite{pukhov} for similar laser pulse parameters, but a wider laser spot ($w \simeq 12.7\lambda=10~\mu$m). The comparison should be made with care because our 2D simulations may overestimate the proton cut-off energy with respect to a realistic 3D case (this numerical effect would be related to the proton acceleration dynamics, since 2D and 3D simulations of electron acceleration in an SP showed similar energies \cite{fedeli2016electron}). However, it is worth noting that Shen et al.\cite{pukhov} may have not observed a cut-off energy increase when shifting the laser pulse (as a test of possible misalignment effects) because the larger value of $w$ makes the shift not sufficient to improve the coupling. The efficient use of a laser pulse with a smaller value of $w$ also suggests that proton energies exceeding 100~MeV might be reached with a lower laser energy than in Ref.\cite{pukhov}, a prediction which needs to be confirmed by future 3D simulations.

The proton energy enhancement at parallel incidence with a shifted laser pulse is absent at a lower intensity of $I=3.4 \times 10^{19}$~W/cm$^2$ (Fig.\ref{Energy_proton_combined}~b): a maximum cut-off energy of $\simeq 25$~MeV is obtained for $\delta=0~\mu$m. 
For grazing incidence, the highest proton energy cut-off of $\simeq 24$~MeV is obtained for $\alpha=1.5^\circ$. All these cases show a less peaked proton spectrum than for the high intensity case of Fig.\ref{Energy_proton_combined}~a).

To understand the mismatch between the maxima of proton and electron energy as a function of $\delta$ for the lower intensity case, we note that the total charge and the density of electrons at the rear edge follow the same trend as the proton energy, but not the same as the electron cut-off energy and temperature. Possible reasons include a beam loading effect, due to the availability of a large electron reservoir in the solid target: the more electrons are accelerated, the more the laser energy is depleted thus quenching the SP and causing saturation in the acceleration process. For the parallel incidence case, at $I=3.4 \times 10^{19}$~W/cm$^2$ the fractional absorption into electrons is $\sim 40\%$ for $\delta=0$ and remains almost constant with increasing shift up to $\delta=2.5~\mu$m before dropping down to $\sim 25\%$: we thus see that the energy depletion of the laser pulse is rather substantial. In contrast, at $I=6 \times 10^{20}$ W/cm$^2$ the fractional absorption is only $\sim 16\%$ for  $\delta=0$ and rises up to a maximum of $\sim 28\%$ for $\delta=3~\mu$m, suggesting that in this case the energy reservoir is large enough to prevent a saturation effect.

 The shift effect we have reported tells us that the target thickness should not be necessarily thinner than the laser waist as in Shen \textit{et al}'s \cite{PhysRevX.11.041002} basic configuration, which relaxes the constraint at least for what concerns electron acceleration. For the optimum shift from the target surface, simulations show the electrons energies remain the same for varying target thickness. The proton cut-off energy remains similar; however, the mono-energetic peak broadens with increasing thicknesses since their number at the target rear edge is proportional to the thickness of the target and if it gets too large the smooth gradient of the sheath field attributed to the large number of electrons in comparison to the protons at the accelerating region is lost.
 
 Recently Marini \textit{et al} \cite{marini2022electron} reported on 3D simulations in a laser-target geometry corresponding to the parallel incidence case $\delta=-d/2$ for semi infinite target, i.e. with the laser hitting on the target wedge. We believe that their observation of electron acceleration is consistent with our theoretical analysis and 2D simulation results.

\section{Conclusion}

We have shown via two-dimensional simulations that efficient laser-driven electron and proton acceleration mediated by surface plasmons may occur at grazing or parallel incidence in simple flat targets, without the need of a grating. For grazing incidence higher values of the electron energy and total charge than in a grating are found for the same laser pulse parameters. The most efficient acceleration is found for parallel incidence when the propagation axis of the laser pulse is shifted by a few wavelengths from the target plane. The investigated regimes can readily be investigated with present-day laser systems and  are very promising for applications. 

\section*{Acknowledgement}
This work was supported by EPSRC (grant EP/p010059/1), and The Ministry of Education, Govt. of India (through University Grants Commission, India) under the collaboration
between Tezpur University, Assam, India and Queens University Belfast,
UK. The EPOCH code was also funded by EPSRC (grants EP/G054950/1, EP/G056803/1, EP/G055165/1 and EP/M022463/1) .We are grateful for use of the computing resources from the Northern Ireland High Performance Computing (NI-HPC) service funded by EPSRC (EP/T022175).

\section*{References}
\bibliographystyle{iopart-num}
\bibliography{bibliography}
\end{document}